\documentclass[a4paper,12pt]{article}
\usepackage{geometry}
\usepackage[cp1251]{inputenc}
\usepackage{epsfig}
\usepackage{graphicx}

\usepackage{amssymb}

\begin{document}

Astronomy Letters, 2016, Vol. 42, No. 5, pp. 314-328 \hspace{20pt} 
Received 01 September 2015

\vspace{-8pt}

\section*{
\begin{center}
T Tauri Stars: Physical Parameters and Evolutionary Status
\end{center}
}

\begin{center}
\Large K. N. Grankin
\end{center}

\begin{center}
\it{Crimean Astrophysical Observatory, Nauchny, Crimea, 298409 Russia}
\end{center}

\begin{center}
konstantin.grankin@mail.ru
\end{center}

\subsection*{\center Abstract}

\begin{quote}

Long-term homogeneous photometry for 35 classical T Tauri stars (cTTS) in
the Taurus–Auriga star-forming region (Tau-Aur SFR) has been analyzed.
Reliable effective temperatures, interstellar extinctions, 
luminosities, radii, masses, and ages have been determined for these 
cTTS. The physical parameters and evolutionary status of 35 cTTS from 
this work and 34 weak-line T Tauri stars (wTTS) from previous studies 
have been compared. The luminosities, radii, and rotation periods of 
low-mass (0.3–1.1~$M_\odot$) cTTS are shown to be, on average, greater 
than those of low-mass wTTS, in good agreement with the evolutionary 
status of these two subgroups. The mean age of the younger subgroup of 
wTTS from our sample (2.3 Myr) essentially coincides with the mean 
duration of the protoplanetary disk accretion phase (2.3 Myr) for a 
representative sample of low-mass stars in seven young stellar 
clusters. The accretion disk dissipation time scale for the younger 
subgroup of cTTS (age \textless\ 4 Myr) in the Tau-Aur SFR is shown to 
be no greater than 0.4 Myr, in good agreement with the short 
protoplanetary disk dissipation time scale that is predicted by 
present-day protoplanetary disk evolution models.\end{quote}

\vspace{15pt}

\textbf{Key words:} {\it stars, pre-main-sequence stars, T Tauri stars 
-- physical properties, evolutionary status.}

\subsection*{INTRODUCTION}
\indent

The Taurus–Auriga complex of molecular clouds is one of the nearest and 
well-studied SFRs. The relatively low interstellar extinction, the 
small distance (about 140 pc), the presence of a rich population of 
young (1–10 Myr) low-mass stars, and the absence of ionizing radiation 
and winds from young massive stars make it an ideal object for studying 
the formation of solar-mass stars and testing theoretical stellar 
evolution models.

By 2008 the population of known young stars in the Tau–Aur SFR had 
numbered 364 objects (Kenyon et al. 2008). According to the empirical 
classification scheme proposed by Lada (1987) and Andr\'e et al. 
(1993), these objects form an approximate age sequence from protostars 
surrounded by thick gas and dust envelopes (Class 0 and I objects) to 
T Tauri stars (TTS) with circumstellar disks (Class II objects) and to 
TTS without accretion disks (Class~III objects). The last two groups 
include cTTS with strong emission lines and significant infrared (IR) 
and ultraviolet (UV) excesses and wTTS with insignificant excesses.
 
The evolution of TTS within the first several million years is 
determined by strong magnetic fields ($\sim$2 kG) and circumstellar 
disks. On the one hand, the magnetic fields and outer deep convective 
envelopes of protostars are responsible for the existence of extended 
cool spots, hot facular fields, excess chromospheric and coronal 
emissions, short flares, and other manifestations of solar-type 
activity that are observed among wTTS. On the other hand, the
magnetic fields play a key role in the complex processes of interaction 
between the central star and its surrounding disk, which lead to a 
redistribution of the angular momentum of the star–disk system and to 
the development of magnetospheric accretion processes typical for cTTS.

Present-day magnetospheric accretion models show that the stellar 
magnetic field acts on the disk at a distance of several stellar radii 
from the stellar surface. The inner disk material ionized by stellar
radiation moves toward the star along magnetic field lines with the 
free-fall velocity, creating hot spots on the stellar surface. Part of 
the ionized gas is ejected back to form a wind and jets (see, e.g., 
Petrov 2003 for a review).
 
The phenomena of solar activity and magnetospheric accretion processes 
produce X-ray, UV, optical, and IR excess emissions that change the shape
of the continuum and veil the photospheric spectral lines. Some of the 
young systems seen in the protoplanetary disk plane can exhibit 
irregular or quasi-periodic fadings attributable to the eclipses of 
much of the stellar surface by the warped inner disk edge, as with 
AA~Tau and related objects (Alencar et al. 2010).

All these phenomena and processes complicate significantly the 
determination of basic physical parameters for young stars (such as the 
luminosity, radius, mass, age, accretion rate, etc.) and their 
evolutionary status, which is very important for testing various models 
for the evolution of young stars.

Attempts to determine the physical parameters of tens of TTS in the 
Tau–Aur SFR have been made in a number of papers (see, e.g., Cohen 
and Kuhi 1979; Strom et al. 1989; Valenti et al. 1993; Kenyon and 
Hartmann 1995; Hartigan et al. 1995; Gullbring et al. 1998; White and 
Ghez 2001; Furlan et al. 2006; Rebull et al. 2010; Furlan et al. 2011;
Andrews et al. 2013; Herczeg and Hillenbrand 2014).

All these studies were performed properly and systematically. 
Nevertheless, the differences in some of the physical parameters for 
the same stars are very large in the above papers. One of the main 
reasons for these differences is the difficulty of estimating the 
interstellar extinction ($A_V$). Hartigan et al. (1995) point out that 
the main source of uncertainty in the $A_V$ estimates is the 
variability of the magnitude, color, and veiling of some TTS. They cite 
the star DR~Tau, whose veiling varies between 6.4 and 20 and whose 
$V$-band photometric variability exceeds one magnitude, as an example. 
Another example is the heavily veiled star DO~Tau, in which the excess 
continuum dominates the photospheric flux. The authors argue that the 
contribution of the veiling continuum in the $V$ band reaches 80\%. 
Thus, the color of this object will be considerably bluer than that of 
the same star without optical veiling. A significant scatter in 
extinction estimates is observed even in the case of TTS with weak 
accretion. For example, for the weakly veiled star DN~Tau, Gullbring et 
al. (1998) provided $A_V\sim0.2$, Kenyon and Hartmann (1995) estimated 
$A_V\sim0.5$, and Hartigan et al. (1995) obtained $A_V\sim1.1$.

Ingleby et al. (2013) provide published extinction estimates for 13 
cTTS and point out that the values of $A_V$ obtained by different 
authors for the same star differ significantly. For half of the objects 
from Ingleby et al. (2013), the scatter in $A_V$ reaches $\pm0.5$, which
can lead to an uncertainty of one order of magnitude in estimating the 
accretion rate and some other physical parameters of cTTS.

To attempt to minimize the uncertainties in the spectral type, 
extinction, and veiling of cTTS, one can apply the method of 
simultaneously fitting all three parameters. Recently, such 
simultaneous fitting has been applied for cTTS in the Tau–Aur
SFR (Fischer et al. 2011; McClure et al. 2013; Herczeg and Hillenbrand 
2014). Nevertheless, the scatter in $A_V$ estimates not only did not 
decrease but even increased to $\pm0.8$. For the three objects DF~Tau,
DO~Tau, and DG~Tau, the differences in $A_V$ estimates reached 
$\pm1.2$, $\pm1.6$, and $\pm2.7$, respectively. Herczeg and Hillenbrand 
(2014) rightly note that the significant uncertainties in $A_V$ and 
other parameters of cTTS introduce scepticism in our ability to use
the main properties of cTTS to test theories of star formation and 
pre-main-sequence evolution.

Quite realistic estimates of $A_V$ and other parameters of TTS can be 
obtained if long-term homogeneous multicolor photometry is used for 
these objects. Analysis of a large number of magnitude estimates in 
several bands (for example, in $U$, $B$, $V$, and $R$) allows one to 
determine the maximum and minimum brightnesses with a high accuracy and 
to establish reliable color–magnitude relations. In turn, knowledge of 
these parameters of the photometric behavior for TTS makes it possible 
to accurately estimate the color excesses that are attributable either 
to accretion processes, or to cool extended spots, or to variable 
circumstellar extinction. Allowance for these excesses enables one to 
determine the magnitudes and colors that correspond to the intrinsic 
TTS photosphere and, hence, to calculate reliable values of $A_V$ and 
other parameters of young stars.

Such long-term homogeneous photometric observations of 34 cTTS and 40 
wTTS in the Tau–Aur SFR had been performed as part of the ROTOR program 
at the Maidanak Observatory in Uzbekistan for almost twenty years (1984–2006).
The results of these observations and their statistical analysis were 
presented in Grankin et al. (2007, 2008). Subsequently, Grankin (2013a) 
analyzed long-term observations for 28 wTTS and 60 wTTS candidates in 
the Tau–Aur SFR. This analysis showed that more than 60 objects from 
this sample exhibit periodic light variations attributable to spotted
rotational modulation. To minimize the influence of photometric 
variability on the $A_V$ estimate and the luminosity $L_{bol}$, we used 
the maximum brightness ($V_{max}$) and the corresponding $(V-R)$ color, which
is virtually insensitive to the possible presence of hot spots and the 
manifestations of chromospheric activity (see Gullbring et al. 1998). 
We hypothesized that the visible wTTS surface was least covered with
spots at the time of maximum light. Therefore, its brightness and color 
correspond most closely to a true photosphere. Because of this, we 
calculated reliable extinctions, luminosities, radii, masses, and ages 
for 74 wTTS and related objects in the Tau–Aur SFR. Based on these 
data, we refined the evolutionary status of these objects (Grankin et 
al. 2013b) and investigated the relationship between activity and
rotation (Grankin 2013c).
 
This paper is a logical continuation of our studies begun in Grankin 
(2013a–2013c). Our goals are: (1) to obtain reliable estimates of the 
basic physical parameters for 35 cTTS in the Tau–Aur SFR based on 
published homogeneous long-term photometric data (Grankin et al. 2007); 
(2) to compare the physical parameters of cTTS and wTTS; and (3) to 
refine their evolutionary status. 
 
\subsection*{DETERMINATION OF STELLAR PROPERTIES}
\indent

The estimates of the luminosity, radius, mass, age, and other stellar 
parameters of cTTS depend primarily on how accurately the effective 
temperature ($T_{eff}$) and interstellar extinction ($A_V$) have been determined.

\subsubsection*{\it Effective Temperature}

The sample presented in this paper includes 35 cTTS from the Tau–Aur 
SFR. Homogeneous long-term photometric data (Grankin et al. 2007) are 
available for the overwhelming majority of objects. Information about 
the cTTS spectral types was taken mainly from Furlan et al. (2011) and, 
only in a few cases, from Fischer et al. (2011), Nguyen et al. (2009), 
G\"udel et al. (2007), and Johns-Krull et al. (2000). These spectral 
types were converted to temperatures using the temperature calibration
from Tokunaga (2000). The arguments for choosing this calibration and 
the corresponding uncertainties in $T_{eff}$ were discussed in detail 
in Grankin (2013a). Note only that before choosing Tokunaga’s 
calibration, we analyzed several different temperature scales from 
Cohen and Kuhi (1979), de Jager and Nieuwenhuijzen (1987), Bessell 
(1991), Kenyon and Hartmann (1995), Tokunaga (2000), Luhman et al. 
(2003), and Herczeg and Hillenbrand (2014). The temperature scales were 
compared with the experimental data on the temperatures and spectral 
types of 43 main-sequence dwarfs from Torres et al. (2010). Since these 
stars are components of noninteracting eclipsing systems, their basic 
physical parameters are known with an accuracy of at least $\pm3\%$. A 
statistical analysis showed that the experimental data from Torres et 
al. (2010) agree best with the temperature calibration from Tokunaga (2000)
with a standard deviation of $\pm90$~K. The temperature calibrations 
mentioned above are presented in Table 1.

If we use the temperature calibration from Tokunaga (2000) and take the 
uncertainty in the spectral classification as $\pm1$ subclass, then the 
corresponding uncertainty in $T_{eff}$ will be $\pm50$, $\pm100$, 
$\pm195$, $\pm90$, and $\pm160$ K for G1–G6, G7–K1, K2–K6, K7–M0,
and M1–M6 stars, respectively. The last row in Table 1 provides the rms 
deviation that characterizes the scatter between the experimental data 
from Torres et al. (2010) and the temperature calibration corresponding 
to a given column. The values calculated by spline interpolation are italicized.

\subsubsection*{\it Interstellar Extinction}

As has been noted in the Introduction, cTTS have considerable X-ray, 
UV, optical, and IR excesses that change the shape of the continuum and make
it more difficult to calculate the interstellar extinction $A_V$ and, 
hence, the bolometric luminosity $L_{bol}$ and all of the remaining 
stellar parameters. On the one hand, the $U-B$ and $B-V$ colors become 
considerably bluer in cTTS due to the presence of hot facular fields, 
excess chromospheric emission, short flares, and hot spots located at 
the base of accretion columns. On the other hand, the colors become 
slightly redder due to the presence of extended cool spots, variable 
circumstellar extinction, additional warm disk emission, and, in some 
cases, due to partial eclipses of the stellar surface by the warped
inner disk edge. Therefore, it is highly problematic to determine the 
magnitudes and colors corresponding to the cTTS photosphere (see, e.g., 
Petrov and Kozak 2007). However, the actual value of $A_V$ cannot be 
estimated without knowing these quantities. To illustrate the 
aforesaid, we presented the $V-R$ color – $V$ magnitude relation in 
Fig. 1 for one wTTS (V827~Tau) and one cTTS (DL~Tau). Both objects have 
the same spectral type K7 and $T_{eff}$ = 4040 K. The photometric data 
are represented by the gray dots. The $(V-R)_{\rm o}$ color of a
standard star of the corresponding spectral type K7 is marked by the 
vertical line.

In the case of V827~Tau (Fig. 1a), the main source of its photometric 
brightness and color variability is the presence of dark photospheric 
spots. Model calculations show that the observed color–magnitude 
relation is best described by dark spots with a temperature lower than 
the photospheric temperature by 1000 K that cover about 67\% of the 
entire stellar surface. The modeling results are represented by the
dash–dotted line. A detailed description of the model calculations can 
be found in Grankin (1998). Since the visible surface of V827~Tau at 
maximum light is least covered with spots, its maximum brightness and
the corresponding color marked by the large white circle correspond 
most closely to a true photosphere. The dashed line indicates the 
brightness level that corresponds to a true stellar photosphere 
observed through absorbing interstellar clouds ($V^\prime_{ph}$). It 
is these values that are used to calculate the interstellar reddening 
$A_V$ (black arrow) and to determine the dereddened magnitude 
$V_{ph}^{\rm o}$ (black circle).

%---------------------------------------------------------------
%\clearpage
%\begin{table}[t]
\begin{table*}
\centering
{{\bf Table 1.} Temperature calibrations}
\label{meansp} 
\vspace{5mm}
%\begin{scriptsize}
\tabcolsep=3.0mm
\begin{small}
\begin{tabular}{r|l|c|c|c|c|c|c} \hline \hline 

\rule{0pt}{2pt}&&&&&&&\\
Sp. T. & CK79$^a$ & B91$^b$ & deJN87$^c$ & KH95$^d$ & L03$^e$ & 
HH14$^f$ & T00$^g$ \\ [5pt]
 \hline
G0      & 5902  & 6000  & 5943  & 6030  &       & 5930  & 5930 \\
G1      & 5834  &\textit{5917} &  & 5945 &  &  & \textit{5876} \\
G2      & 5768  & \textit{5833} & 5794 & 5860 & & 5690  & 5830 \\
G3      &       &       &       & 5830  &    & & \textit{578}6 \\
G4      &       &       & 5636  & 5800  &       &       & 5740 \\
G5&5662& \textit{5583}&\textit{5554}&5770& &5430&\textit{5687} \\
G6      & 5445  & 5500  &       & 5700  &       &       & 5620 \\
G7      &       &       &       & 5630  &    & & \textit{5535} \\
G8      &       &      & 5309 & 5520  & & 5180 & \textit{5438} \\
G9      &       &       &       & 5410  &    & & \textit{5337} \\
K0      & 5236  & \textit{5167}  & 5152 & 5250 & & 4870 & 5240 \\
K1&5105 &\textit{5083} &4989 & 5080 & &\textit{4790}&\textit{5145} \\
K2      & 4955  & 5000  & \textit{4808}  & 4900  & & 4710 & 5010 \\
K3&4775 &\textit{4750} &4688 & 4730 & & \textit{4543}&\textit{4801} \\
K4      & 4581  & 4500  & 4539  & 4590  & &\textit{4377} & 4560 \\
K5      & 4395  &       & 4406  & 4350  &       & 4210  & 4340 \\
K6      & 4198  &       &       & 4205  &    & & \textit{4166} \\
K7      & 3999  & 4000  & 4150  & 4060  &       & 4020  & 4040 \\
K8      &       &       &       &       &    & & \textit{3954} \\
K9      &       &       & 3936  &       &    & & \textit{3883} \\
M0      & 3917  & 3800  & 3837  & 3850  &       & 3900  & 3800 \\
M1      & 3681  & 3650  & 3664  & 3720  & 3705  & 3720  & 3680 \\
M2      & 3499  & 3500  & 3524  & 3580  & 3560  & 3560  & 3530 \\
M3      & 3357  & 3350  & 3404  & 3470  & 3415  & 3410  & 3380 \\
M4      & 3228  & 3150  & 3289  & 3370  & 3270  & 3190  & 3180 \\
M5      & 3119  & 3000  & 3170  & 3240  & 3125  & 2980  & 3030 \\
M6      & 2958  & 2800  & 3034  & 3050  & 2990  & 2860  & 2850 \\
\hline
SD&$\pm$99K&$\pm$111K&$\pm$136K&$\pm$134K&$\pm$140K&$\pm$242K&$\pm$
90K \\
\hline

\multicolumn{8}{l}{}\\ [-3mm]
\multicolumn{8}{l}{\footnotesize \textbf{Notes.} a -- Cohen and Kuhi, 1979;
b -- Bessell, 1991; c -- de Jager and Nieuwenhuijzen, 1987;}\\
\multicolumn{8}{l}{\footnotesize d -- Kenyon and Hartmann, 1995; e -- 
Luhman et al., 2003; f -- Herczeg and Hillenbrand, 2014;}\\
\multicolumn{8}{l}{\footnotesize g -- Tokunaga, 2000.}\\

\end{tabular}
\end{small}
\end{table*}
%---------------------------------------------------------------

In the case of the cTTS DL~Tau (Fig. 1b), the brightness and color 
variations are attributable primarily to the presence of a hot spot at 
the base of an accretion column. For this reason, the minimum 
brightness and the corresponding color (white circle), when the hot 
spot is located on the invisible side of the star, will correspond (to 
a first approximation) to a true photosphere. Our modeling shows that the
observed color–magnitude relation corresponds most closely to a hot 
spot with a temperature of $\sim10000$ K and an area of $\sim1.7\%$ of 
the entire stellar surface (dash–dotted line). Obviously, in the case 
of cTTS, not the maximum brightness and the corresponding color (as in 
the case of wTTS) but the minimum brightness and the corresponding 
color should be used to obtain realistic interstellar extinction 
estimates (black arrow). However, the source of the cTTS photometric 
variability is not only the hot accretion spot but also variable 
circumstellar extinction (as in AA~Tau), dark spots, hot facular 
fields, and some other processes (see above). Therefore, the choice of 
the brightness ($V^\prime_{ph}$) and the corresponding color that 
correspond most closely to a “quiet” cTTS photosphere and that could be 
used to determine a realistic value of $A_V$ is highly ambiguous. 
Below, we present the next algorithm to determine $A_V$.

%---------------------------------------------------------------
\begin{figure}[ht]
\epsfxsize=15cm
\vspace{0.6cm}
\hspace{1cm}\epsffile{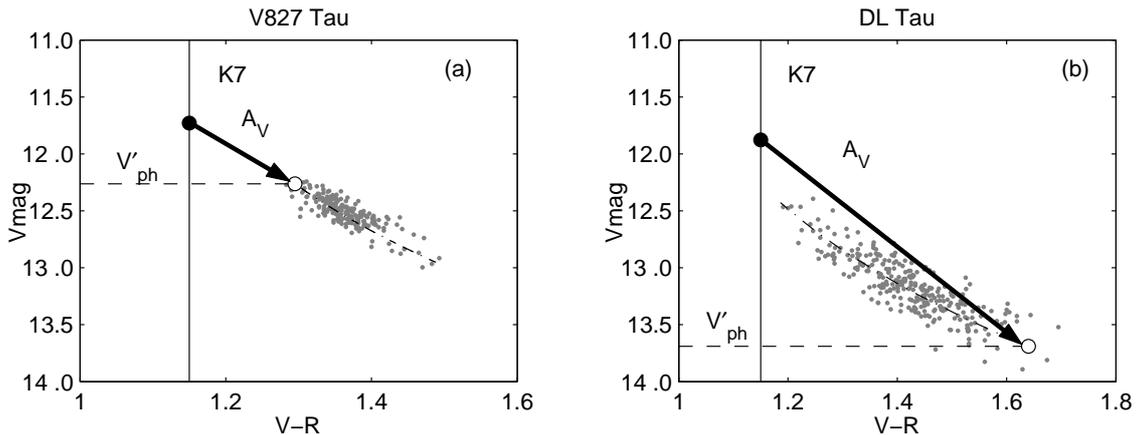}
\caption{\rm \footnotesize {$V-R$ color – $V$ magnitude relation for 
wTTS (a) and cTTS (b). The photometric data are marked by the gray 
dots. The solid vertical line indicates the color of a standard dwarf 
star of the same spectral type K7 without interstellar extinction. The
large white circle corresponds to the presumed magnitude 
$V^\prime_{ph}$ and color $(V-R)_{ph}$ that refer to the “quiet” 
stellar photosphere. The dashed horizontal line indicates the 
brightness level $V^\prime_{ph}$. The large black circle indicates the 
dereddened magnitude and color of the star. The black arrow indicates 
the interstellar reddening $A_V$. The modeling results that describe 
best the photometric behavior of the objects are represented by the 
dash–dotted line. The observed color–magnitude relation for the wTTS 
V827~Tau corresponds to the model with dark spots with a temperature 
lower than the photospheric temperature by 1000 K and an area of 67\% 
of the entire stellar surface. In contrast, the photometric behavior of 
the cTTS DL~Tau can be explained by the presence of a hot spot with a 
temperature of $\sim10000$ K and an area of $\sim1.7\%$ of the entire 
stellar surface.}}
\end{figure}

To take into account the possible contribution of the emission from a 
hot spot and facular areas, we used long-term homogeneous $BVR$ photometry
from Grankin et al. (2007). Analysis of the color–magnitude diagrams 
showed that all cTTS from our sample exhibit considerable blue excesses 
(estimates of the color excess $E_{U-B}$, $E_{B-V}<0$). Figure 2a 
presents the $B-V$ color – $V$ magnitude relation for the star CI~Tau, 
which shows signatures of noticeable accretion. The $(B-V)_{\rm o}$ 
color of a standard star of the corresponding spectral type K7 is 
marked by the vertical line. The actual color–magnitude relation is
marked by the dash–dotted line. It can be seen from the figure that 
many of the magnitudes and colors are located to the left of the 
vertical line, where $E_{B-V}<0$. Obviously, these magnitudes and $B-V$ 
colors are attributable to the accretion processes, and they cannot be 
used to estimate $A_V$. Nevertheless, we can assume that the brightness 
level at which the blue excess disappears ($E_{B-V}=0$) corresponds to 
a normal stellar photosphere observed through interstellar clouds. This 
brightness level ($V^\prime_{ph}$) is marked by the horizontal dashed 
line and the white circle.

Given the upper photospheric brightness limit $V^\prime_{ph}$ and the 
$V-R$ color–magnitude relation, we can estimate the color excess $E_{V-R}$
attributable only to the interstellar extinction (Fig. 2b). We propose to
use such a value that corresponds to $V^\prime_{ph}$ (marked by the 
vertical dotted line) as the photospheric $(V-R)_{ph}$ color. The 
photospheric brightness and the corresponding $(V-R)_{ph}$ color chosen 
in this way are marked by the white circle. According to the proposed 
technique, the color excess due to the interstellar extinction is 
estimated as $E_{V-R}=0.39$. Finally, the interstellar extinction was 
calculated from the well-known formula $A_V=3.7E_{V-R}$ for the $V$ and $R$
bands of the Johnson (1968) photometric system.

%---------------------------------------------------------------
\begin{figure}[ht]
\epsfxsize=15cm
\vspace{0.6cm}
\hspace{1cm}\epsffile{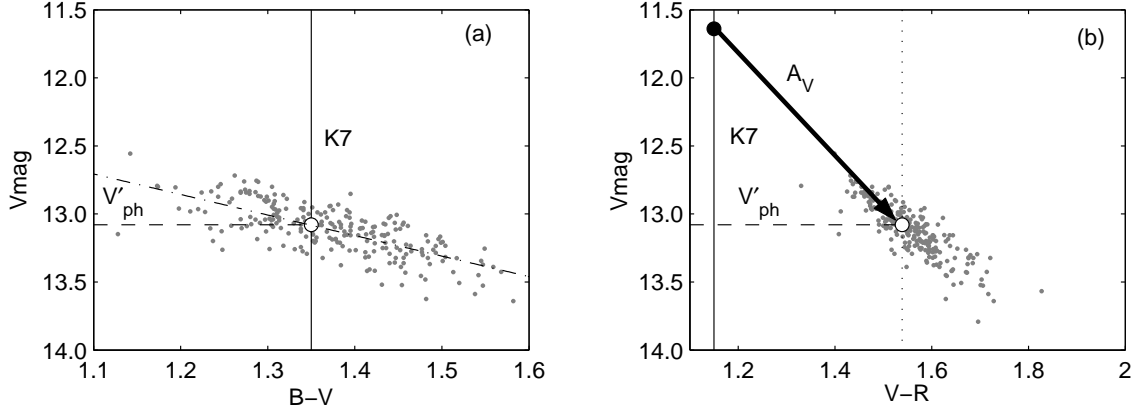}
\caption{\rm \footnotesize {(a) $B-V$ color – magnitude relation for 
the cTTS CI~Tau. The designations are the same as those in Fig. 1. The 
gray dots represent the photometric data. The solid vertical line 
indicates the $B-V$ color of a standard dwarf star of the same spectral 
type as that of the cTTS under study but without interstellar 
extinction. The dash–dotted line is a linear fit to the photometric data.
The dashed horizontal line indicates the brightness level of the cTTS 
under study ($V^\prime_{ph}$) for which the color excess $E_{B-V} = 0$. 
(b) $V-R$ Color–magnitude relation for the cTTS CI~Tau. The solid 
vertical line indicates the $V-R$ color of a standard dwarf star of the 
same spectral type as that of the cTTS under study but without 
interstellar extinction. The dotted vertical line indicates the 
$(V-R)_{ph}$ color of the cTTS under study that corresponds to 
$V^\prime_{ph}$. The large white circle corresponds to the magnitude 
$V^\prime_{ph}$ and $(V-R)_{ph}$ color that are free from the accretion 
excess and that are used to calculate the interstellar extinction $A_V$ 
(black arrow). The large black circle indicates the dereddened 
magnitude and color of the star.}}
\end{figure}
%---------------------------------------------------------------

To check the reliability of our extinction estimates, we compared them 
with the published values of $A_V$. Information regarding the published 
values of $A_V$ was taken from 18 different papers, with the references 
to most of them being given in the Introduction. Figure 3 compares our 
values of the extinction with those from several publications with which
good agreement is observed. Our values of $A_V$ agree best with the 
published estimates from Cohen and Kuhi (1979), Strom et al. (1989), 
and Kenyon and Hartmann (1995). These authors used the $V-R$ or $V-I$ 
colors to measure the extinction. As an example, our values of $A_V$ 
are compared with those from Strom et al. (1989) in Fig. 3a. There is an
excellent overlap between our values and the results of these authors, 
with a shift of $-0.08^m$ and a standard deviation of $0.48^m$. The 
agreement with the data from Cohen and Kuhi (1979) (a shift of 
$-0.01^m$ and a standard deviation of $0.57^m$) and from Kenyon and
Hartmann (1995) (a shift of $-0.06^m$ and a standard deviation of 
$0.57^m$) is slightly poorer.

The differences with the extinction estimates from Rebull et al. (2010) 
and Furlan et al. (2011) obtained from near-IR photometry ($J-H$ or 
$J-K$) are more significant and systematic. For example, the mean shift 
and standard deviation of our values of AV relative to those from 
Rebull et al. (2010) are $+1.25^m$ and $3.75^m$, respectively. When our 
values of $A_V$ are compared with those from Furlan et al. (2011), the
situation is slightly better: the mean shift is $+0.59^m$ and the 
standard deviation is $1.08^m$ (see Fig. 3b). In both cases, analysis 
of the near-IR photometry gives considerably higher extinction 
estimates than those obtained at optical wavelengths.

%---------------------------------------------------------------
\begin{figure}[ht]
\epsfxsize=14cm
\vspace{0.6cm}
\hspace{1cm}\epsffile{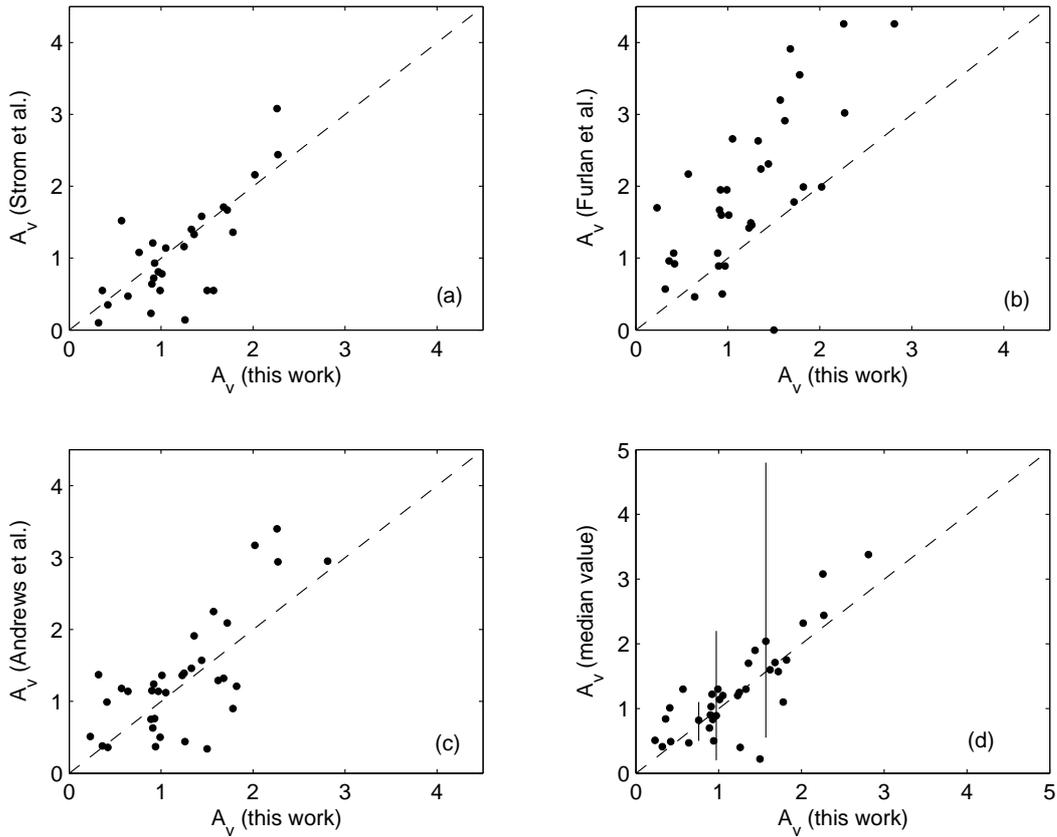}
\caption{\rm \footnotesize {Comparison of $A_V$ from this paper with 
its values from Strom et al. (1989) (a), Furlan et al. (2011) (b), Andrews
et al. (2013) (c), and the median values of $A_V$ (d). The three vertical
bars indicate a typical scatter between the maximum and minimum values of
$A_V$ for three different objects: DI~Tau ($\Delta A_V=0.60^m$), DE~Tau 
($\Delta A_V=2.0^m$), and RY~Tau ($\Delta A_V=4.3^m$). Despite the large
differences between individual published $A_V$ estimates, the median 
values agree excellently with our extinction estimates, with a shift of 
$+0.08^m$ and a standard deviation of $0.18^m$.}}
\end{figure}
%---------------------------------------------------------------

Andrews et al. (2013) applied a method for estimating the extinction 
and luminosity by fitting spectral templates of stellar photosphere 
models to the actual spectral energy distributions of cTTS. Our values 
of $A_V$ are compared with those from Andrews et al. (2013) in Fig. 3c. 
The agreement with our data is good, with a shift of $+0.17^m$ and a 
standard deviation of $0.31^m$. Unfortunately, comparison with other 
works showed very significant discrepancies in the $A_V$ estimates. For 
example, analysis of 12 works showed that the maximum and minimum 
values of $A_V$ for CW~Tau differ by $5.1^m$ (from 1.8 to 6.9); the 
differences for DG~Tau, RY~Tau, DO~Tau, and DD~Tau reach 4.4, 4.3, 4.2, 
and $3.9^m$, respectively. For most of the remaining objects from our
sample, the differences between the maximum and minimum values of $A_V$ 
are close to $2^m$. Taking this fact into account, we gathered all of 
the independent $A_V$ determinations from the literature for our objects
and calculated the median values of the extinction. Figure 3d compares 
our extinction estimates with these median values of $A_V$. The 
vertical bars indicate a typical scatter between the maximum and 
minimum values of $A_V$ for three different objects: DI~Tau 
($\Delta A_V=0.60^m$), DE~Tau ($\Delta A_V=2.0^m$), and RY~Tau 
($\Delta A_V=4.3^m$). Despite the large differences between individual 
published $A_V$ estimates, the median values agree excellently with our 
extinction estimates, with a shift of $+0.08^m$ and a standard 
deviation of $0.18^m$. Thus, we think that our technique provides a 
good way for estimating reliable values of the interstellar extinction.

\subsubsection*{\it Luminosity and Radius}

The bolometric luminosity ($L_{bol}$) was calculated using a well-known 
formula: $\log (L_*/L_\odot) = -0.4(V_{max} -A_V +BC+5 -5\log r-4.72)$, 
where $BC$ is the bolometric correction from Hartigan et al. (1994), 
and $r$ is the mean distance to the Tau–Aur SFR (140 pc). The various 
sources of errors in the $L_{bol}$ estimates were discussed in Grankin 
(2013a). The most significant errors can be caused by the possible 
presence of a secondary component and the uncertainty in the adopted 
distance to the Tau–Aur SFR.

About 70 multiple systems in the Tau–Aur SFR are known to date (Harris 
et al. 2012). In our sample consisting of 35 cTTS, 13 objects are 
components of binary or multiple systems. The magnitude difference for 
the components of three objects is more than $1.8^m$, and the errors in 
$L_{bol}$ are insignificant for them. For the remaining 10 objects, we 
can overestimate $L_{bol}$ in the worst case by a factor of 2 if the
magnitudes of the components are assumed to be the same.

An equally serious problem can be associated with the uncertainty in 
the distance. According to several accurate individual trigonometric 
parallaxes (an error of $\sim0.4\%$) obtained through VLBI measurements
(Loinard et al. 2007; Torres et al. 2009, 2012), the mean distance to 
the Tau–Aur SFR is $d = 140.6\pm 13.6$ pc. This value is in good 
agreement with its previous estimates ($d = 140\pm 20$ pc). Thus, an
uncertainty in the distance of $\pm 20$ pc can lead to an error in 
$\log L_{bol}$ of the order of $\pm0.13$ dex.

The stellar radii were determined by two methods. First, we estimated 
the radii ($R_{bol}$) using $T_{eff}$ and $L_{bol}$. Second, we used 
the ratio from Kervella and Fouqu\'e (2008). These authors obtained an 
empirical relation between the angular diameters of the nearest
dwarfs and their visible colors. They showed that the angular diameter 
could be calculated with an accuracy of at least $5\%$. We calculated 
the stellar radii ($R_{KF}$) by using the Kervella–Fouqu\'e calibration 
and by assuming the mean distance to the Tau–Aur SFR to be 140 pc. 
$R_{bol}$ is plotted against $R_{KF}$ in Fig. 4. The estimates of the 
radii are in good agreement with the mean ratio
$<R_{KF}/R_{bol}> = 0.964 \pm0.017$.

\begin{figure}[ht]
\epsfxsize=10cm
\vspace{0.6cm}
\hspace{4cm}\epsffile{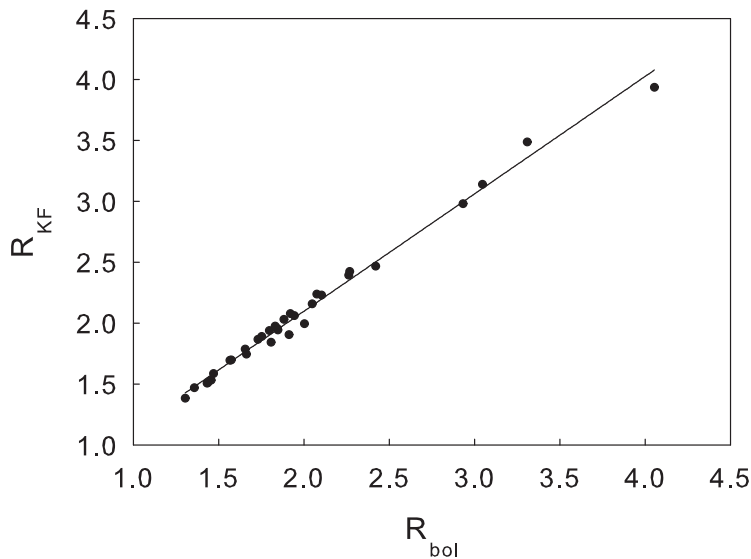}
\caption{\rm \footnotesize {Comparison of the stellar radii $R_{bol}$ 
obtained using $L_{bol}$ and $T_{eff}$ with the radii $R_{KF}$ 
calculated from the Kervella–Fouqu\'e relation. The estimates of the 
radii are in good agreement with the mean ratio 
$<R_{KF}/R_{bol}> = 0.964 \pm0.017$.}}
\end{figure}

\subsubsection*{\it Mass and Age}

To determine the masses and ages of the stars from our sample, we used 
the grid of evolutionary tracks from Siess et al. (2000) computed for 
pre-mainsequence (PMS) stars. The Hertzsprung–Russell (HR) diagram for 
35 cTTS is presented in Fig. 5a. For comparison, Fig. 5b shows the HR 
diagram for 17 well-known wTTS (filled squares) and 17 new wTTS (open 
squares) from Grankin (2013b). The inaccuracies in the mass and age 
estimates are attributable to the uncertainties in $T_{eff}$ and 
$L_{bol}$ adopted in this paper and depend on the object’s position on
the HR diagram. The error in the mass is $\pm0.1~M_\odot$ for the stars 
on convective tracks and $\pm0.2~M_\odot$ for the objects on radiative 
tracks. The relative error in the age is of the order of $\pm1-4$ Myr.

\begin{figure}[ht]
\epsfxsize=8.5cm
\vspace{0.6cm}
\hspace{4cm}\epsffile{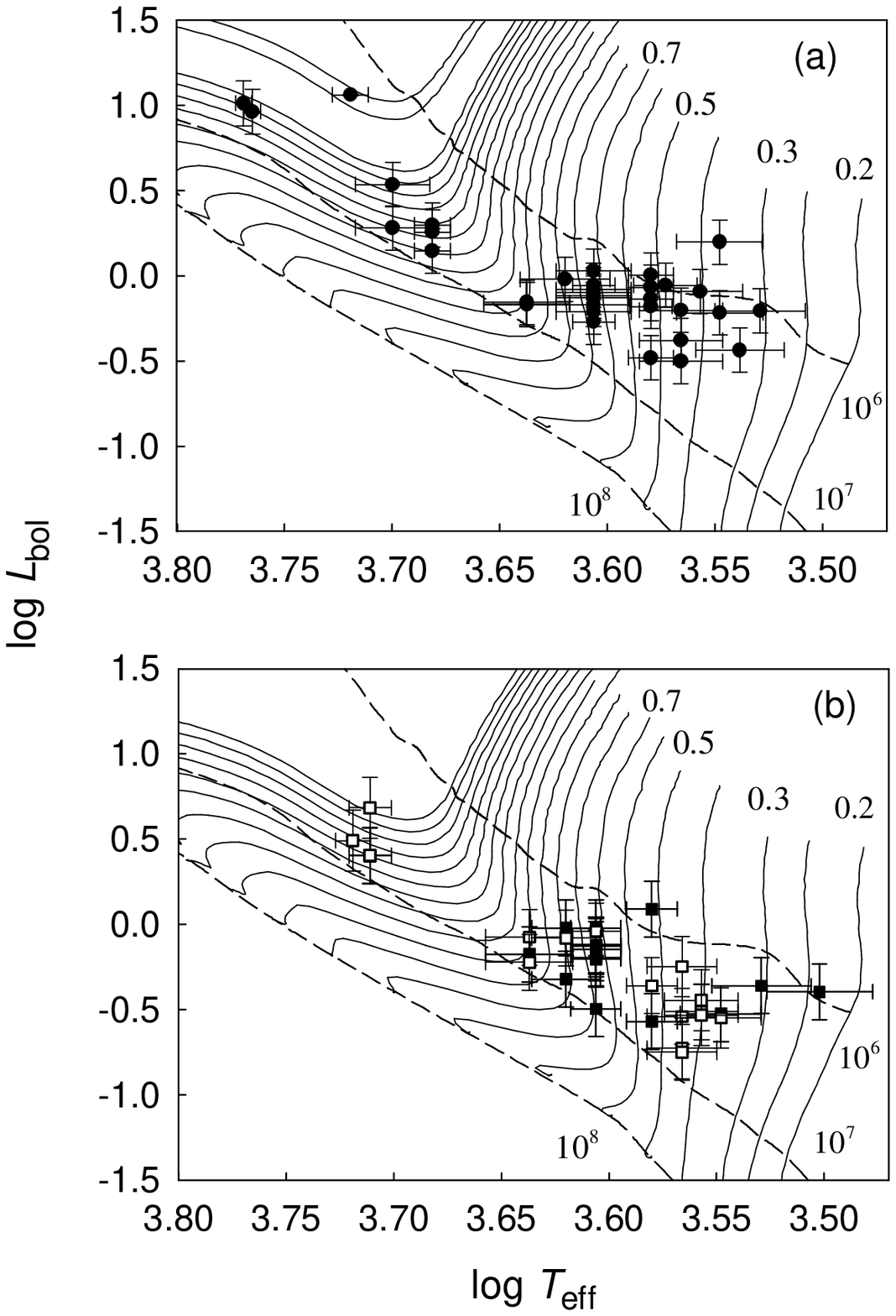}
\caption{\rm \footnotesize {HR diagram for cTTS (a) and wTTS (b). The 
filled circles are cTTS, the filled squares are well-known wTTS, and
the open squares are new wTTS. The errors indicate the $\pm1\sigma$ 
uncertainty for $L_{bol}$ and $T_{eff}$. The solid lines are the evolutionary
tracks calculated with Y = 0.277 and Z = 0.02 for stars with masses of 0.2,
0.25, 0.3, 0.4, 0.5, 0.6, 0.7, 0.8, 0.9, 1.0, 1.1, 1.2, 1.3, 1.4, 1.5, 1.6,
1.7, 1.8, 1.9, 2.0, 2.5, and 2.7~$M_\odot$. The dashed lines are the 
isochrones for ages of $10^6$, $10^7$, and $10^8$ yr.}}
\end{figure}

It can be seen from Fig. 5 that the overwhelming majority of cTTS and wTTS
occupy the same region on the HR diagram that corresponds to the range of
masses from 0.3 to 1.1~$M_\odot$. Only three wTTS and seven cTTS are on the
evolutionary tracks corresponding to more massive PMS stars with masses 
from 1.4 to 2.0~$M_\odot$. The star T Tau, the TTS prototype, is at the 
very top of the HR diagram between the tracks that correspond to PMS 
stars with masses of 2.5 and 2.7~$M_\odot$.

The basic cTTS parameters and the intermediate data used to calculate 
them are presented in Table 2. Since the same photometric database and 
the same technique were used to determine the physical parameters of 35 
cTTS from this paper and 34 wTTS from Grankin (2013b), the systematic 
errors in these parameters must be minimal for the two samples of young 
stars. Therefore, comparing the parameters and evolutionary status of 
cTTS and wTTS from the Tau–Aur SFR is of considerable interest.

\subsection*{EVOLUTIONARY STATUS}
\indent

The main difference between the cTTS and wTTS is that the cTTS have 
accretion disks, while the wTTS are devoid of such disks or at least their
inner regions. For this reason, the wTTS must rotate faster and be 
slightly older than the cTTS. If this assumption is valid and the wTTS 
are actually older objects, then their luminosities and radii must be
smaller than those of the cTTS. To test this assumption, we constructed 
the corresponding histograms.

Figure 6 shows the distributions (histograms) of luminosities and radii 
for both cTTS (Figs. 6a and 6c) and wTTS (Figs. 6b and 6d). In both 
samples of stars, there are several objects with luminosities 
$>1.5L_\odot$. If we neglect these objects and select the best 
Gaussians for stars with luminosities in the range from 0.18 to 
1.23~$L_\odot$, then we will find that the mean cTTS luminosity 
(0.70$L_\odot\pm0.20 L_\odot$) is slightly greater than the mean wTTS 
luminosity (0.56$L_\odot\pm0.27 L_\odot$). Accordingly, for the 
overwhelming majority of objects with radii in the range from 1 to 
2.5~$R_\odot$, we find that the mean cTTS radius 
(1.83$R_\odot\pm0.29R_\odot$) is slightly larger than the mean wTTS 
radius (1.59$R_\odot\pm0.30R_\odot$).

The distributions (histograms) of ages and masses for the cTTS and wTTS 
are presented in Fig. 7. It follows from the figure that the age and 
mass distributions for both subgroups have a complex shape, and they 
cannot be classified as simple unimodal (in this case, they must have 
had one pronounced peak) or uniform (in this case, the histograms would 
contain an approximately equal number of values in each bin). The age 
and mass distributions (histograms) presented in Fig. 7 are closest in 
their shape to bimodal ones. Unfortunately, we can say nothing about 
the significance of the bimodal shape of the distributions based on a 
small data sample (35 cTTS and 34 wTTS). Reliably determining the shape 
of a complex distribution, including the bimodal one, requires a 
considerably larger volume of input data: from 400 to 2000 values (see, 
e.g., Novitskii and Zograf 1985). Nevertheless, analysis of Figs. 7a 
and 7b shows that the age distribution has a bimodal structure for both
subgroups: young stars with ages of 1–4 Myr and objects with ages of 
5–10 Myr.

The approximation using two Gaussians gives two maximum values of about 
$1.9\pm0.9$ and $6.1\pm1.3$ Myr for the cTTS and $2.3\pm0.8$ and 
$7.0\pm1.5$ Myr for the wTTS. Thus, the mean age of the older cTTS 
subgroup is smaller than the mean age of the analogous wTTS subgroup by 
0.9 Myr. At the same time, the mean age of the younger cTTS subgroup is smaller
than the mean age of the analogous wTTS subgroup by 0.4 Myr. Since the 
difference between the mean ages of the younger cTTS and wTTS subgroups 
is smaller than the relative error in the age (see the previous section),
we estimated the significance of this difference based on Student’s 
t-test or, more specifically, the significance was calculated from the
confidence interval and the smallest significant difference (SSD). Both 
methods showed the difference between the mean ages of the younger cTTS 
and wTTS subgroups to be statistically insignificant at the 90\% 
confidence level.

%---------------------------------------------------------------
\begin{table}[t]

\vspace{6mm}
\centering
{{\bf Table 2.}  Basic cTTS parameters and the intermediate data used 
to calculate them. The masses and ages were estimated in comparison 
with the theoretical models and tracks calculated by Siess et al. (2000)}

\vspace{5mm}

\begin{scriptsize}
\begin{tabular}{r|l|c|c|c|c|c|c|c|c|c|c|} \hline \hline 
\rule{0pt}{2pt}&&&&&&&&&&&\\
Name & Sp. type & Ref. & $T_{eff}$, K & $V^\prime_{ph}$ &$V-R$ & $E_{V-R}$
& $A_V$ & $ L_{bol}$, $L_\odot$ & $R, R_\odot$ & $M, M_\odot$ & $t, 10^6$
yr \\ [5pt]
\hline 
    AA Tau &   K7 & 1 & 4040 & 12.94 & 1.40 & 0.25 & 0.92 & 0.54 & 1.47 & 0.77 & 2.09 \\
    BP Tau &   K7 & 1 & 4040 & 12.38 & 1.39 & 0.24 & 0.89 & 0.88 & 1.88 & 0.75 & 3.16 \\
    CI Tau &   K7 & 1 & 4040 & 13.08 & 1.54 & 0.39 & 1.44 & 0.76 & 1.75 & 0.74 & 2.49 \\
    CW Tau &   K3 & 1 & 4801 & 12.14 & 1.35 & 0.53 & 1.95 & 1.85 & 1.94 & 1.50 & 4.90 \\
    CX Tau & M2.5 & 1 & 3455 & 13.72 & 1.65 & 0.10 & 0.36 & 0.37 & 1.66 & 0.34 & 1.66 \\
    CY Tau &   M1 & 2 & 3680 & 13.66 & 1.74 & 0.34 & 1.26 & 0.63 & 1.92 & 0.45 & 1.31 \\
    DD Tau & M3.5 & 1 & 3280 & 14.83 & 2.13 & 0.48 & 1.78 & 0.62 & 2.26 & 0.27 & 0.29 \\
    DE Tau & M1.5 & 1 & 3605 & 13.23 & 1.71 & 0.26 & 0.97 & 0.81 & 2.27 & 0.41 & 0.99 \\
    DF Tau &   M2 & 1 & 3530 & 12.65 & 1.77 & 0.27 & 0.99 & 1.58 & 3.31 & 0.37 & 0.13 \\
  DG Tau A &   K6 & 1 & 4166 & 12.88 & 1.51 & 0.44 & 1.62 & 0.95 & 1.85 & 0.87 & 2.41 \\
    DH Tau &   M1 & 1 & 3680 & 13.73 & 1.56 & 0.16 & 0.57 & 0.31 & 1.36 & 0.45 & 2.92 \\
    DI Tau &   M0 & 3 & 3800 & 12.84 & 1.49 & 0.21 & 0.76 & 0.66 & 1.84 & 0.53 & 1.52 \\
    DK Tau &   K7 & 1 & 4040 & 12.88 & 1.51 & 0.36 & 1.33 & 0.83 & 1.83 & 0.74 & 2.22 \\
    DL Tau &   K7 & 1 & 4040 & 13.69 & 1.64 & 0.49 & 1.82 & 0.61 & 1.57 & 0.75 & 3.39 \\
    DM Tau &   M1 & 1 & 3680 & 14.35 & 1.81 & 0.41 & 1.50 & 0.42 & 1.57 & 0.45 & 2.01 \\
    DN Tau &   M0 & 1 & 3800 & 12.38 & 1.39 & 0.11 & 0.42 & 0.73 & 1.94 & 0.53 & 1.35 \\
  DO Tau E $^{a}$ &   M0 & 1 & 3800 &       &      &      & 2.27 & 1.01 & 2.25 & 0.52 & 0.97 \\
    DR Tau &   K5 & 1 & 4340 & 12.72 & 1.32 & 0.33 & 1.23 & 0.68 & 1.43 & 1.10 & 7.06 \\
    DS Tau &   K5 & 1 & 4340 & 12.42 & 1.24 & 0.25 & 0.93 & 0.68 & 1.44 & 1.10 & 7.06 \\
  GG Tau A &   K7 & 4 & 4040 & 12.18 & 1.40 & 0.25 & 0.91 & 1.07 & 2.08 & 0.72 & 1.62 \\
    GI Tau &   K7 & 1 & 4040 & 13.03 & 1.52 & 0.37 & 1.36 & 0.74 & 1.73 & 0.74 & 2.57 \\
    GK Tau &   K7 & 1 & 4040 & 12.59 & 1.42 & 0.27 & 1.01 & 0.80 & 1.80 & 0.74 & 2.33 \\
    GM Aur &   K7 & 1 & 4040 & 12.08 & 1.24 & 0.09 & 0.32 & 0.68 & 1.66 & 0.75 & 2.89 \\
 HP Tau AB$^{b}$ &   K3 & 3 & 4801 &       &      &      & 2.26 & 1.40 & 1.77 & 1.39 & 6.90 \\
    HQ Tau &   K2 & 1 & 5010 & 12.23 & 1.50 & 0.76 & 2.81 & 3.43 & 2.42 & 1.88 & 4.02 \\
    IP Tau &   M0 & 1 & 3800 & 13.05 & 1.34 & 0.06 & 0.23 & 0.33 & 1.30 & 0.54 & 3.77 \\
    IQ Tau $^{b}$& M0.5 & 3 & 3740 &       &      &      & 1.25 & 0.88 & 2.20 & 0.51 & 1.06 \\
    RW Aur &   K3 & 1 & 4801 & 11.16 & 1.06 & 0.24 & 0.87 & 1.69 & 1.85 & 1.50 & 6.09 \\
    RY Tau &   G1 & 1 & 5876 &  9.55 & 0.94 & 0.42 & 1.57 &10.28 & 3.05 & 1.93 & 5.93 \\
    SU Aur &   G2 & 4 & 5830 &  9.01 & 0.77 & 0.24 & 0.90 & 9.18 & 2.93 & 1.88 & 6.40 \\
     T Tau &   K0 & 1 & 5240 &  9.83 & 1.11 & 0.47 & 1.72 &11.51 & 4.05 & 2.64 & 2.06 \\
  UX Tau A &   K2 & 5 & 5010 & 10.68 & 0.91 & 0.17 & 0.64 & 1.91 & 1.81 & 1.55 & 8.62 \\
    UY Aur &   M0 & 1 & 3800 & 12.84 & 1.56 & 0.28 & 1.05 & 0.86 & 2.10 & 0.52 & 1.13 \\
 V1079 Tau &   K5 & 1 & 4340 & 11.87 & 1.10 & 0.11 & 0.41 & 0.70 & 1.46 & 1.10 & 6.78 \\
    XZ Tau &   M2 & 1 & 3530 & 14.39 & 1.95 & 0.45 & 1.68 & 0.61 & 2.05 & 0.37 & 1.19 \\ 
\hline
\multicolumn{12}{l}{}\\ [-3mm]
\multicolumn{12}{l}{\scriptsize \textbf{} 1 -- Furlan et al. 2011;
2 -- Fischer et al. 2011; 3 -- G\"udel et al. 2007; 4 -- Nguyen et al. 2009;
5 -- Johns-Krull et al. 2000.}\\
\multicolumn{12}{l}{\scriptsize \textbf{} a -- $A_V$ and $L_{bol}$
from Gullbring et al. 1998; b -- $A_V$ and $L_{bol}$ from G\"udel et al.
2007.}\\

\end{tabular} 
\end{scriptsize}
\end{table}
%---------------------------------------------------------------}
\clearpage

\begin{figure}[ht]
\epsfxsize=9.9cm
\vspace{0.0cm}
\hspace{2.5cm}\epsffile{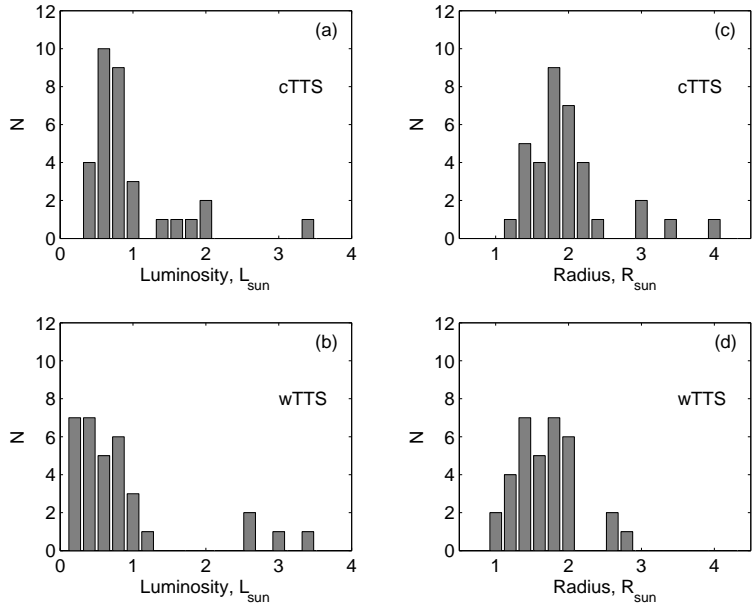}
\caption{\rm \footnotesize {Distributions of luminosities (a, b) and 
radii (c, d) for cTTS and wTTS.}}
\end{figure}

\begin{figure}[ht]
\epsfxsize=9.9cm
\vspace{0.0cm}
\hspace{2.5cm}\epsffile{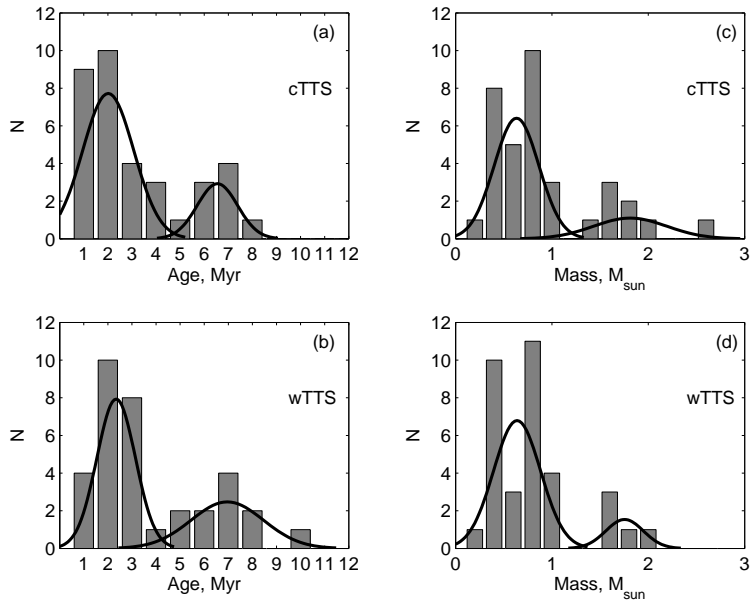}
\caption{\rm \footnotesize {Distributions of ages (a, b) and masses 
(c, d) for cTTS and wTTS.}}
\end{figure}

%\clearpage

It may well be that the bimodal pattern of the age distribution is 
attributable to the selection effects in our data. A more 
representative sample of wTTS and cTTS should be analyzed. If the age 
distribution is actually bimodal, then there is reason to assert that
the star formation process in the Tau–Aur SFR was more likely cyclic 
than continuous.

The approximation using two Gaussians for the cTTS and wTTS mass 
distributions allows two subgroups to be identified: a subgroup of 
stars with masses of 0.2–1.1~$M_\odot$ and a subgroup of stars with
masses of 1.4–2.6~$M_\odot$ (see Figs. 7c and 7d, respectively). The 
mean masses of these two subgroups are $0.63\pm0.23$ and 
$1.81\pm0.38~M_\odot$ for the cTTS and $0.64\pm0.23$ and 
$1.75\pm0.19~M_\odot$ for the wTTS. It can be argued that the mass 
distributions for the cTTS and wTTS essentially coincide.

It has been noted above that we cannot determine the significance of 
the bimodal shape of the age and mass distributions due to the limited 
size of our sample. Nevertheless, our data on the mass distribution are 
consistent with the results of studying the initial mass function (IMF) 
in the Tau–Aur SFR. In contrast to other well-studied star-forming regions,
the IMF for the Tau-Aur SFR has an unusual excess of stars with masses 
of 0.6–0.8~$M_\odot$ and a clear deficit of objects with masses 
noticeably greater than 1~$M_\odot$ (see, e.g., Luhman et al. 2009). 
Recent hydrodynamic simulations showed that the peculiar shape of the 
IMF could be a direct consequence of the unusual properties of the 
cores from which small groups of protostars were formed in the Tau–Aur SFR
(Goodwin et al. 2004). Roughly 50\% of the young stars formed in a core 
are ejected from the core to form a population of low-mass stars and 
brown dwarfs with a flat IMF. The remaining objects form multiple 
systems within the core, gradually accrete matter, and produce a 
population of intermediate-mass stars whose IMF peaks at $\sim0.6–0.8~M_\odot$.

To compare the rotation of cTTS and wTTS, we selected objects with 
known rotation periods (Artemenko et al. 2012; Grankin 2013a), ages 
$<\ $4~Myr, and masses in the range 0.2–1.1~$M_\odot$. As a result, the
cTTS and wTTS samples contain 17 and 15 objects, respectively. The 
distributions of cTTS and wTTS rotation periods are presented in 
Figs.~8a and 8b, respectively. It follows from the figure that most of
the cTTS (12 of 17) have rotation periods in the range from 5.5 to 8.1 
days, three objects rotate with periods of 3.2–4.6 days, and only two 
stars have periods longer than 10 days. In contrast, among the wTTS 
there are only four objects with periods in the range from 5.5 to 8.1 
days, and 60\% of the stars have rotation periods in the range from 0.6 
to 3.8 days. A statistical analysis showed that the mean cTTS and wTTS 
rotation periods are $6.98\pm2.73$ and $4.31\pm2.56$ days, respectively.
The 95\% confidence intervals for these mean values coincide and are
equal to 1.3 days.

Our estimate of the mean age for the younger and larger wTTS subgroup 
(2.3 Myr) agrees excellently with the time at which the disk accretion phase
ceases (2.3 Myr) from Fedele et al. (2010). These authors performed an 
optical spectroscopic survey of a representative sample of low-mass 
(K0–M5) stars from seven young stellar clusters (with ages from 2 to
30 Myr) and showed the disk accretion phase to cease in the 
overwhelming majority of these stars at an age of 2.3 Myr. This result 
is in good agreement with the present views of the evolutionary status 
of wTTS that lost their original disks through the accretion and 
formation of planetary systems.

The second result of our study suggests that the mean age of the 
younger and larger cTTS subgroup (1.9 Myr) is smaller than the mean age 
of the analogous wTTS subgroup in the Tau–Aur SFR by 0.4 Myr. Since all 
cTTS from our sample show signatures of accretion disks, while the wTTS 
lost their disks, it should be concluded that the accretion disks in 
the Tau–Aur SFR dissipate within a fairly short time interval, 
$\sim$0.4 Myr. This result agrees well with the previous studies of the 
dissipation time scale for protoplanetary disks. They showed that
the dissipation time scale for the original disk measured from the time 
at which the accretion process ceases does not exceed 0.5 Myr (see, 
e.g., Skrutskie et al. 1990; Wolk and Walter 1996; Cieza et al. 2007).
The disk evolution models combining viscous accretion with 
photoevaporation successfully reproduce the fairly long lifetime of the 
original disk and the short disk dissipation time scale after the 
cessation of accretion (see, e.g., Alexander et al. 2006a, 2006b). A
more detailed discussion of the evolution of a protoplanetary disk and 
the “two-time-scale” problem can be found in the review of Williams and 
Cieza (2011).

Comparison of the wTTS and cTTS rotation periods showed that the CTTS 
rotate, on average, more slowly (6.98 days) than do the wTTS (4.31 days).
This result is consistent with the theoretical models that predict a 
decrease in the rotation period of PMS stars as they move toward the 
zero-age main sequence. Previous studies of the evolution of angular 
momentum for PMS stars in the Tau–Aur SFR showed that most of the 
objects ceased to actively interact with their disks on a time scale from
0.7 to 10 Myr. Spin-up due to a rapid decrease in the moment of inertia 
is observed after disk dissipation (see, e.g., Grankin 2013b). Since 
the cTTS continue to actively interact with their accretion disks, they
evolve with an almost constant angular velocity, and their rotation 
periods are grouped in a narrow range, from 5.5 to 8.1 days. In 
contrast, the wTTS lost their disks and rotate freely, remaining under 
the action of only the braking wind that carries away some part of
the angular momentum. Therefore, the wTTS rotate faster, and many of 
them have rotation periods in the range from 0.6 to 3.8 days.

\begin{figure}[h]
\epsfxsize=7.0cm
\vspace{0.6cm}
\hspace{5.0cm}\epsffile{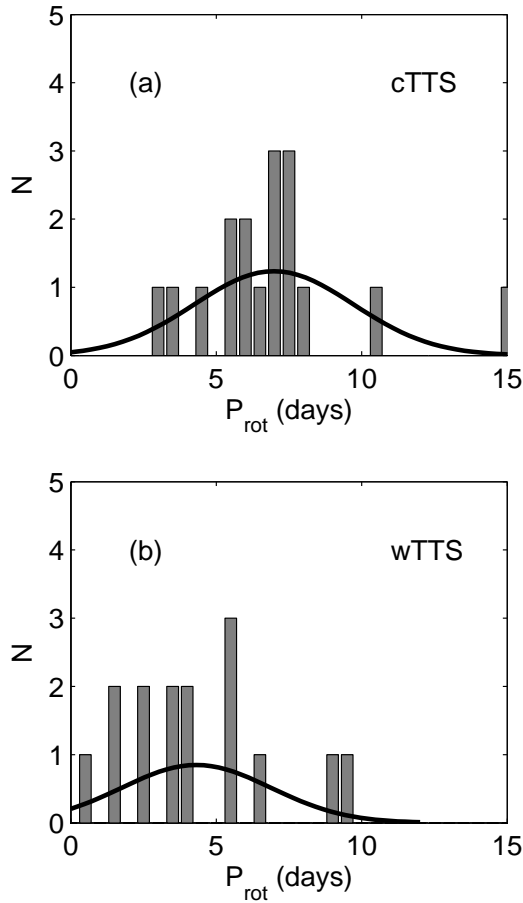}
\caption{\rm \footnotesize {Distributions of rotation periods for cTTS 
(a) and wTTS (b).}}
\end{figure}

\subsection*{CONCLUSIONS}
\indent

Based on published homogeneous long-term photometric data (Grankin et 
al. 2007), we calculated reliable effective temperatures, interstellar 
extinctions, luminosities, radii, masses, and ages for 35 cTTS in the 
Tau–Aur SFR. $T_{eff}$ were determined using the temperature
calibration from Tokunaga (2000). The errors in $T_{eff}$ can reach 
$\pm50$, $\pm100$, $\pm195$, $\pm90$, and $\pm160$ K for G1–G6, G7–K1, 
K2–K6, K7–M0, and M1–M6 stars, respectively.

Our analysis of the homogeneous long-term $BVR$ photometry allowed the 
brightness level of each cTTS at which the blue excess attributable to 
the accretion processes disappears ($E_{B-V}=0$) to be estimated. 
This brightness level ($V^\prime_{ph}$) corresponds best to a normal 
stellar photosphere. To calculate the interstellar extinction $A_V$ , 
we used the $(V-R)_{ph}$ color that corresponded to $V^\prime_{ph}$. 
These extinction estimates were compared with those from 18 different
papers. Our values of $A_V$ were shown to agree best with the published 
estimates from Cohen and Kuhi (1979), Strom et al. (1989), and Kenyon and
Hartmann (1995). In contrast, significant systematic differences are 
observed for $A_V$ from Rebull et al. (2010) and Furlan et al. (2011), 
where the extinction was calculated using near-IR photometry. Our 
extinction estimates were shown to agree excellently with the median 
values of $A_V$ calculated using all the published values of this quantity.

We calculated the stellar bolometric luminosity $L_{bol}$ by assuming 
all cTTS to be at the mean distance of the Taurus–Auriga SFR (140 pc), 
except for T~Tau itself and HP~Tau, whose more accurate distances are 
known from special studies. We showed that the actual scatter of 
distances to the Tau–Aur SFR ($\pm20$ pc) could lead to an error in 
$\log L_{bol}$ of the order of $\pm0.13$ dex.

The stellar radii were calculated by two methods: using $L_{bol}$ and 
$T_{eff}$ ($R_{bol}$) and using the Kervella–Fouqu\'e ratio ($R_{KF}$). 
The $R_{bol}$ and $R_{KF}$ estimates were shown to be in excellent 
agreement with the mean ratio $<R_{KF}/R_{bol}> = 0.964 \pm0.017$.

The masses and ages were determined using a grid of evolutionary tracks 
from Siess et al. (2000). The error in the mass for the stars on 
convective and radiative tracks is $\pm0.1~M_\odot$ and about $\pm0.2~M_\odot$,
respectively. The relative error in the age is of the order of $\pm 1-4$
Myr. The overwhelming majority of cTTS (28 of 35 objects) have low 
masses in the range from 0.3 to 1.1~$M_\odot$.

We compared the physical parameters and evolutionary status of 35 cTTS 
from this study and 34 wTTS with a reliable evolutionary status from
Grankin (2013b). We identified two groups of objects among the wTTS and 
cTTS: low-mass (0.2–1.1~$M_\odot$) stars and stars with masses 
1.4–2.6~$M_\odot$. The low-mass cTTS were shown to have, on average, a 
high luminosity and a large radius than do the low-mass wTTS. In 
addition, the low-mass cTTS rotate, on average, more slowly 
($<P_{rot}>=6.98$ days) than do the low-mass wTTS ($<P_{rot}>=4.31$ days).
These results are in good agreement with the evolutionary status of the 
investigated objects.

Our analysis of the age distribution of stars showed that the cTTS and 
wTTS exhibit the same bimodal distribution: there are stars with ages 
1–4 Myr and objects with ages 5–10 Myr. The mean age of the older cTTS 
subgroup (6.1 Myr) is smaller than the mean age of the analogous wTTS 
subgroup (7.0 Myr) by 0.9 Myr, while the mean age of the younger and 
more representative cTTS subgroup (1.9 Myr) is smaller than the mean 
age of the analogous wTTS subgroup (2.3 Myr) by 0.4 Myr. It may well be 
that the bimodal pattern of the age distribution is attributable to the 
selection effects.

The mean age of the younger subgroup of wTTS from our sample (2.3 Myr) 
essentially coincides with the mean duration of the accretion phase 
(2.3 Myr) determined by analyzing a representative sample of low-mass 
stars in the young stellar clusters $\sigma$ Orionis, NGC~6231, 
NGC~6531, ASCC~58, NGC~2353, Collinder~65, and NGC~6664 (Fedele et al. 
2010). This result is consistent with the present views of the 
evolutionary status of wTTS that lost their original disks through the 
accretion and formation of planetary systems.

The dissipation time scale of accretion disks in the Tau–Aur SFR was 
shown to be no greater than 0.4 Myr, in good agreement with the previous
estimates of the protoplanetary disk dissipation time scale (see, e.g., 
Williams and Cieza 2011 for a review).

%%%%%%%%%%%%%%%%%%%%%%
\subsection*{REFERENCES}

\begin{enumerate}

\item S.H.P. Alencar, P.S. Teixeira, M.M. Guimar$\tilde a$es, et al.,
Astron. Astrophys. {\bf 519}, A88 (2010).
\vspace{-1ex}
\item R.D. Alexander, C.J. Clarke, and J.E. Pringle, Mon.
Not. R. Astron. Soc. {\bf 369}, 216 (2006a).
\vspace{-1ex}
\item R.D. Alexander, C.J. Clarke, and J.E. Pringle, Mon.
Not. R. Astron. Soc. {\bf 369}, 229 (2006b).
\vspace{-1ex}
\item P. Andr\'e, D. Ward-Thompson, and M. Barsony, Astrophys. J. 
{\bf 406}, 122 (1993).
\vspace{-1ex}
\item S.M. Andrews, K.A. Rosenfeld, A.L. Kraus, et al., Astrophys. J. 
{\bf 771}, 129 (2013).
\vspace{-1ex}
\item S.A. Artemenko, K.N. Grankin, and P.P. Petrov,
Astron. Lett {\bf 38}, 783 (2012).
\vspace{-1ex}
\item M. Bessell, Astron. J. {\bf 101}, 662 (1991).
\vspace{-1ex}
\item L. Cieza, D.L. Padgett, K.R. Stapelfeldt, et al.,
Astrophys. J. {\bf 667}, 308 (2007).
\vspace{-1ex}
\item M. Cohen and L. Kuhi, Astrophys. J. Suppl. Ser. {\bf 41},
743 (1979).
\vspace{-1ex}
\item D. Fedele, M.E. van den Ancker, T. Henning, et al.,
Astron. Astrophys. {\bf 510}, 72 (2010).
\vspace{-1ex}
\item W. Fischer, S. Edwards, L. Hillenbrand, et al.,
Astrophys. J. {\bf 730}, 73 (2011).
\vspace{-1ex}
\item E. Furlan, L. Hartmann, N. Calvet, et al., 
Astrophys. J. Suppl. Ser. {\bf 165}, 568 (2006).
\vspace{-1ex}
\item E. Furlan, K. L. Luhman, C. Espaillat, et al.,
Astrophys. J. Suppl. Ser. {\bf 195}, 3 (2011).
\vspace{-1ex}
\item S.P. Goodwin, A.P. Whitworth, and D. Ward-Thompson, Astron. 
Astrophys. {\bf 419}, 543 (2004).
\vspace{-1ex}
\item K.N. Grankin, Astron. Lett. {\bf 24}, 497 (1998).
\vspace{-1ex}
\item K.N. Grankin, S.Yu. Melnikov, J. Bouvier, et al., Astron. 
Astrophys. {\bf 461}, 183 (2007).
\vspace{-1ex}
\item K.N. Grankin, J. Bouvier, W. Herbst, et al., Astron. Astrophys. 
{\bf 479}, 827 (2008).
\vspace{-1ex}
\item K.N. Grankin, Astron. Lett. {\bf 39}, 251 (2013a).
\vspace{-1ex}
\item K.N. Grankin, Astron. Lett. {\bf 39}, 336 (2013b).
\vspace{-1ex}
\item K.N. Grankin, Astron. Lett. {\bf 39}, 446 (2013c).
\vspace{-1ex}
\item M. G\"udel, K.R. Briggs, K. Arzner, et al., Astron. Astrophys. 
{\bf 468}, 353 (2007).
\vspace{-1ex}
\item E. Gullbring, L. Hartmann, C. Brice\~no, et al.,
Astrophys. J. {\bf 492}, 323 (1998).
\vspace{-1ex}
\item R.J. Harris, S.M. Andrews, D.J. Wilner, et al., 
Astrophys. J. {\bf 751}, 115 (2012).
\vspace{-1ex}
\item P. Hartigan, S. Edwards, and L. Ghandour, Astrophys. J. 
{\bf 452}, 736 (1995).
\vspace{-1ex}
\item G.J. Herczeg and L.A. Hillenbrand, Astrophys. J. 
{\bf 786}, 97 (2014).
\vspace{-1ex}
\item L. Ingleby, N. Calvet, G. Herczeg, et al., Astrophys. J. 
{\bf 767}, 112 (2013).
\vspace{-1ex}
\item C. de Jager and H. Nieuwenhuijzen, Astron. Astrophys. 
{\bf 177}, 217 (1987).
\vspace{-1ex}
\item C.M. Johns-Krull, J.A. Valenti, and J.L. Linsky,
Astrophys. J. {\bf 539}, 815 (2000).
\vspace{-1ex}
\item H.L. Johnson, in {\it Nebulae and Interstellar Matter},
Ed. by B.M. Middlehurst and L.H. Aller (Univ.
Chicago Press, Chicago, 1968), p. 167.
\vspace{-1ex}
\item S.J. Kenyon and L. Hartmann, Astrophys. J. Suppl. Ser. 
{\bf 101}, 117 (1995).
\vspace{-1ex}
\item S.J. Kenyon, M. G\'omez, and B.A. Whitney, in {\it Handbook of 
Star Forming Regions}, Vol. 1: {\it The Northern Sky}, Ed. by Bo 
Reipurth (ASP Monograph, 2008), Vol. 4, p. 405.
\vspace{-1ex}
\item P. Kervella and P. Fouqu\'e, Astron. Astrophys. 
{\bf 491}, 855 (2008).
\vspace{-1ex}
\item C.J. Lada, in {\it Star Forming Regions, Proceedings
of the IAU Symp.} No. 115, Ed. by M. Peimbert
and J. Jugaku (Cambridge Univ. Press, Cambridge,
1987), p. 1.
\vspace{-1ex}
\item L. Loinard, R.M. Torres, A.J. Mioduszewski, et al., 
Astrophys. J. {\bf 671}, 546 (2007).
\vspace{-1ex}
\item K.L. Luhman, J.R. Stauffer, A.A. Muench, et al.,
Astrophys. J. {\bf 593}, 1093 (2003).
\vspace{-1ex}
\item K.L. Luhman, E.E. Mamajek, P.R. Allen, et al., 
Astrophys. J. {\bf 703}, 399 (2009).
\vspace{-1ex}
\item M.K. McClure, N. Calvet, C. Espaillat, et al., 
Astrophys. J. {\bf 769}, 73 (2013).
\vspace{-1ex}
\item D.C. Nguyen, R. Jayawardhana, M.H. van Kerkwijk, et al.,
Astrophys. {\bf 695}, 1648 (2009).
\vspace{-4ex}
\item P.V. Novitskii and I.A. Zograf, {\it Estimation of Measurement 
Errors} (Energoatomizdat, Leningrad, 1985), p. 248 [in Russian].
\vspace{-1ex}
\item P.P. Petrov, Astrophysics {\bf 46}, 506 (2003).
\vspace{-1ex}
\item P.P. Petrov and B.S. Kozak, Astron. Rep. {\bf 51}, 500 (2007).
\vspace{-1ex}
\item L.M. Rebull, D.L. Padgett, C.-E. McCabe, et al., 
Astrophys. J. Suppl. Ser. {\bf 186}, 259 (2010).
\vspace{-1ex}
\item L. Siess, E. Dufour, and M. Forestini, Astron. Astrophys. 
{\bf 358}, 593 (2000).
\vspace{-1ex}
\item M.F. Skrutskie, D. Dutkevitch, S.E. Strom, et al., 
Astron. J. {\bf 99}, 1187 (1990).
\vspace{-1ex}
\item K.M. Strom, F.P. Wilkin, S.E. Strom, et al., 
Astron. J. {\bf 98}, 1444 (1989).
\vspace{-1ex}
\item A. Tokunaga, {\it Allen’s Astrophysical Quantities}, 4th
ed., Ed. by A. N. Cox (Springer, New York, 2000), p. 143.
\vspace{-1ex}
\item R.M. Torres, L. Loinard, A.J. Mioduszewski, et al., 
Astrophys. J. {\bf 698}, 242 (2009).
\vspace{-1ex}
\item G. Torres, J. Andersen, and A. Gimenez, 
Astron. Astrophys. Rev. {\bf 18}, 67 (2010).
\vspace{-1ex}
\item R.M. Torres, L. Loinard, A.J. Mioduszewski, et al.,
Astrophys. J. {\bf 747}, 18 (2012).
\vspace{-1ex}
\item J.A. Valenti, G. Basri, and C.M. Johns, 
Astron. J. {\bf 106}, 2024 (1993).
\vspace{-1ex}
\item R.J. White and A.M. Ghez, Astrophys. J. {\bf 556}, 265 (2001).
\vspace{-1ex}
\item J.P. Williams and L.A. Cieza, Ann. Rev. Astron.
Astrophys. {\bf 49}, 67 (2011).
\vspace{-1ex}
\item S.J. Wolk and F.M. Walter, Astron. J. {\bf 111}, 2066 (1996).

\end{enumerate}

\end{document}